\documentstyle[12pt]{article}
\def\psfig#1{}

\def\Tr{\mathop{\rm Tr}}
\def\w{w}

\begin{document}
\renewcommand{\thefootnote}{\fnsymbol{footnote}}

\begin{center}
{\Large {\bf Spin/disorder correlations and \\ duality in the $c = 1/2$ 
string}}\footnote[1]{This work was supported in part by 
the National Science Foundation under grants PHY/9200687 and PHY/9108311,
by the U.S. Department of Energy (D.O.E.) under cooperative agreement 
DE-FC02-94ER40818, by the divisions of 
applied mathematics of the D.O.E. under contracts DE-FG02-88ER25065
and DE-FG02-88ER25065, and by the European Community Human Capital Mobility 
programme.}\\
\vspace{0.2in}
Sean M. Carroll,$^{(1)}$ Miguel E. Ortiz$^{(2)}$\footnote[2]{Address after
January 1996: Blackett Laboratory, Imperial College, Prince Consort Road,
London SW7 2BZ, UK.} and 
Washington Taylor IV$^{(1)}$\footnote[3]{Address after January 1996:
Department of Physics, Joseph Henry Laboratories, Princeton
University, Princeton NJ 08544.}\\
\vspace{0.1in}
{\it $^{(1)}$Center for Theoretical Physics, Laboratory for
Nuclear Science}\\
{\it and Department of Physics}\\
{\it Massachusetts Institute of Technology}\\
{\it Cambridge, Massachusetts\quad 02139}\\
{\it email: carroll@ctp.mit.edu, wati@mit.edu}\\
\vspace{0.1in}
{\it $^{(2)}$Institute of Cosmology, Department of Physics and Astronomy,}\\
{\it Tufts University, Medford, MA 02155}\\
{\it and}\\
{\it Instituut voor Theoretische Fysika, Universiteit Utrecht}\\
{\it Princetonplein 5, 3508 TA Utrecht, The Netherlands}\\
{\it email: ortiz@fys.ruu.nl}\\
\end{center}
\begin{abstract}
We use the method of discrete loop equations to calculate exact correlation
functions of spin and disorder operators on the sphere and on the
boundary of a disk in the $c = 1/2$ string, both in the Ising and dual
Ising matrix model formulations.  For both the Ising and dual Ising
theories the results on the sphere are in agreement with the KPZ/DDK
scaling predictions based on Liouville theory; the results on the
disk agree with the scaling predictions of Martinec, Moore, and
Seiberg for boundary operators.  The calculation of Ising
disorder correlations on the sphere requires the use of boundary variables
which have no matrix model analog.
A subtlety in the calculation on the disk arises because the
expansions of the correlation functions have leading singular terms
which are nonuniversal; we show that this issue may be resolved
by using separate cosmological constants for each boundary domain.
These results give evidence that the Kramers-Wannier duality symmetry of 
the $c = 1/2$ conformal field theory survives
coupling to quantum gravity, implying a duality symmetry of the $c =
1/2$ string even in the presence of boundary operators.
\end{abstract}

\renewcommand{\thefootnote}{\arabic{footnote}}
\vfill
\centerline{CTP~\# 2481 \hfill October 1995}
\centerline{THU-95/29\hfill}
\centerline{hep-th/9510208 \hfill}
\eject
\section{Introduction}

Our understanding of 2D Euclidean quantum gravity coupled to conformal 
matter fields has improved dramatically in the last decade,
largely due to the development of matrix model methods.  The connection
between the discretized and continuum formulations of the theory,
however, still involves a number of unresolved issues.   Although
the relationship between bulk operators in the continuum and in matrix 
models has been fairly well established \cite{lz1,lz2,ks2,bk} (see
\cite{dgz} for a review), the  boundary operators in these theories are 
less well understood and have a more subtle relationship 
\cite{mss,mms,kostov2,ks}.  While the matrix model formalism
is naturally suited to the calculation of correlation functions of
boundary operators \cite{mms}, few explicit computations of
such correlation functions between nontrivial matter fields have been
performed (one calculation of this type is described in \cite{ks}).

In this paper we consider the simplest theory with nontrivial matter
fields, the $c = 1/2$ model which describes a single free fermion
coupled to Liouville gravity.  We compute exact correlation
functions of operators in the discretized theory, both on the sphere and 
with operators inserted on
the boundary of a disk.  The $c = 1/2$ theory has two distinct
descriptions as a matrix model, which can be constructed by coupling
Ising spins to a triangulation or its dual graph respectively
\cite{kazakov,boukaz,kostov}.  Although Kramers-Wannier duality relates 
these two formulations of the Ising model in flat space, it is not
clear {\it a priori} that this duality still holds after coupling
to quantum gravity.  In fact, it has been pointed out that there is a
puzzle if one assumes that this duality exists \cite{staudacher}, in
that the leading singular term in the two-disorder-operator
correlation function on the boundary of a disk in the standard $c =
1/2$ matrix model seems to disagree with results for the scaling
dimension of the disorder operator in the dual theory.  We resolve
this puzzle by calculating the disorder operator correlation functions
in both formulations of the theory, and showing that the universal
behavior of these correlation functions is identical in the two
models.  On the disk, the leading singular terms in the two models
agree, but are nonuniversal; the universal terms are determined by using
separate cosmological constants for each boundary domain.  The results
presented here give strong evidence that Kramers-Wannier
duality persists under coupling to quantum gravity.

One particularly interesting feature of the calculation on the sphere
is that, although the scaling behavior of the spin operator in the
usual matrix model formulation of the theory can be determined
by calculating the correlation function of the disorder operator in the 
dual formulation \cite{kostov}, it does not seem to be possible to
calculate the correlation function of the disorder operator (the spin
operator in the dual formulation) by matrix model methods.  To
calculate this correlation function we use the geometric loop
equation methods detailed in \cite{cot1}, taking advantage of the
freedom in that formalism to use different boundary variables than those 
associated with the matrix model description.

In Section 2, we review the matrix model descriptions of the $c = 1/2$
string as an Ising model coupled to dynamically triangulated 2D
gravity, both in the standard and dual formulations.  We 
include a discussion of
the cubic equations satisfied by the homogeneous disk amplitude in
each of these models.  In Section 3, we take the continuum limit of
the homogeneous disk amplitude in both models, and compare to the
predictions of Liouville theory.  In Section 4, we compute the
correlation functions of spin and disorder operators on the sphere in
the Ising and dual models.  In Section 5, we derive the correlation
functions on the disk and compare with Liouville predictions.  Section
6 contains concluding remarks.

While this manuscript was being completed, we received a preprint
by Sugino and Yoneya \cite{sy} containing some results which overlap with 
those presented in Section 5.                    

\section{Discrete $c = 1/2$ models}

There are two ways of describing the $c = 1/2$ string in terms of a
matrix model, each of which has been analyzed extensively in the
literature.  The original formulation of this theory as a matrix model 
is implemented by putting an
Ising spin at the center of each triangle in a dynamically
triangulated gravity theory \cite{kazakov,boukaz}.  The corresponding
matrix model action is given by
\begin{equation}
Z
= \int {\rm D} U \; {\rm D} V \;
\exp \left(-N\left[\frac{1}{2} {\rm Tr}\; U^2 + 
\frac{1}{2}{\rm Tr}\;  V^2 -c{\rm Tr}\; UV -
\frac{g}{3}  ({\rm Tr}\; U^3 +{\rm Tr}\; V^3) \right] \right)\ ,
\label{eq:matrixising}
\end{equation}
where $g$ and $c$ are coupling constants and $U$ and $V$ are $N\times N$
hermitian matrices.  (Note that the coupling
constant $c$ is unrelated to the central charge $c = 1/2$.  An
unfortunate historical coincidence of notation makes it difficult to
change either symbol; in this paper the symbol $c$ always refers to
the coupling constant except when it appears in the equation $c =
1/2$.)  

In the alternate formulation of the $c = 1/2$ theory, Ising spins
are coupled to the vertices of the dynamical triangulation.  This
corresponds to the Kramers-Wannier dual to (\ref{eq:matrixising}), and
may be thought of as a special case of the $O (n)$ model on a random
surface \cite{kostov,duk,ez}.               
The action for this matrix model is
\begin{equation}
Z = \int {\rm D}  X \; {\rm D}  Y \;
\exp \left(-N\left[ \frac{(1-c)}{2} {\rm Tr}\; X^2 + \frac{(1 + c)}{2}
{\rm Tr}\; Y^2 
-\frac{\widehat{g}}{3} ({\rm Tr}\; X^3 + 3 {\rm Tr}\; XY^2)
\right] \right).
\label{eq:matrixdual}
\end{equation}
In this model, edges of the dual graph connecting two $Y$ matrices
(the $Y$ propagators) correspond to boundaries of constant spin
domains in the dual Ising theory.  In the remainder of this paper, for
simplicity, we refer to the two models (\ref{eq:matrixising}) and
(\ref{eq:matrixdual}) as the ``Ising theory'' and the ``dual theory''
respectively.

\subsection{Duality}

The two matrix model formulations of the $c = 1/2$ string are ontologically 
quite different; in the Ising theory spins are associated with faces of the
random triangulation, whereas in the dual theory spins are associated 
with vertices of the dual graph of such a triangulation, whose coordination
number is unconstrained.  It is therefore interesting to recognize
that a simple coordinate transformation \cite{ez,johnston,cot1}     
\begin{eqnarray}
 X & \rightarrow &  \displaystyle \frac{1}{ \sqrt{2}}({U + V})
 \nonumber\\
 Y & \rightarrow &  \displaystyle \frac{1}{ \sqrt{2}}({U - V})
 \label{eq:transform}\\
 \widehat{g} &  \rightarrow &  \displaystyle g/{ \sqrt{2}}\nonumber
\end{eqnarray}
reveals that the partition functions for the two models are identical
in the absence of boundaries.  The equality of the partition functions
for these two models, however, does not imply that the
Kramers-Wannier duality of the flat space theory survives the coupling
to quantum gravity.  The transformation (\ref{eq:transform}) has a
nontrivial action both on the operators and the states of the theory,
taking the spin operator $U-V$ in the original theory to the disorder
operator $Y$ in the dual theory, and taking free boundary conditions in
the original theory to fixed spin boundary conditions in the dual theory 
(and vice-versa).  To demonstrate that duality persists after coupling to 
quantum gravity, on the other hand, it is necessary to show that the
correlation functions of the theory are the same in both formulations.
In fact, we will show in this paper that unlike the partition
function, the correlation functions in the two theories are not
identical at the discrete level, but that the universal parts of the
correlation functions which correspond to the continuum limit of these
models {\it are} identical.

\subsection{Solution of models}

Because the action for the Ising theory is of the form
\begin{equation}
S(U,V) = {\rm Tr}\; (f_U (U)+ f_V (V)+ UV)\ ,
\label{eq:simpleaction}
\end{equation}
the matrices $U$ and $V$ can be simultaneously diagonalized
\cite{iz,mehta,kazakov,boukaz}, and the method of orthogonal polynomials 
can be used to solve for the correlation functions
\begin{equation}
p_n \equiv\frac{1}{ N} \left\langle{\rm Tr}\; U^n \right\rangle\ .
\end{equation} 
These correlation functions describe the disk amplitude with a
boundary condition corresponding to all Ising spins being fixed  in
the same direction.
Using loop equation methods it can be shown that the generating
function
\begin{equation}
\phi(u)\equiv{1\over N}\left\langle{\rm Tr}\; {1\over 1-uU}\right\rangle 
= \sum_{k = 0}^{\infty}   u^k p_k
\label{eq:phidef}
\end{equation}
satisfies a cubic equation.  (We evaluate all matrix model expectation
values in the large-$N$ limit.)  This cubic equation has been derived by a
number of authors using different methods
\cite{gn,alfaro,staudacher,dl,cot1} and is given by
\begin{equation}
  f_3 \phi^3 + f_2 \phi^2 +f_1 \phi + f_0 = 0\ ,
  \label{eq:purecubic}
\end{equation}
where
\begin{equation}
  \begin{array}{rcl}
  f_0 &=&   -{g^3}+g^2 \left( 2 - {g}p_1 \right) u  -
   g\left( 1 + c - 2{g}p_1 + {g^2}p_2 \right) 
    {u^2} 
  \\
  &&\qquad -\left( {g^2} + g(1+c)p_1 - 
      2{g^2}p_2 + {g^3}p_3 -c(1-{c^2})\right) {u^3}
  \\
  f_1 &=&  {g^3} -2{g^2}u +g\left(1+c \right) {u^2}  -
   \left( c(1 - {c^2}) + {g^2} \right) {u^3} +
   g\left(1- {g}p_1 \right) {u^4}
  \\
  f_2 &=& u^3(2 g^2 -2 g u + c  u^2)
  \\
  f_3 &=& g u^6\ .
  \end{array}
\end{equation}
The coefficients $p_1,p_2,p_3$ can be determined either from the analyticity
properties of this cubic, or by the method of orthogonal polynomials.
$p_1$ and $p_3$ are given as functions of $c$ and $g$ by the relations
\begin{equation}
p_1= {\beta^2 \left[\left(1-2\beta\right)^2
\left(-1 + 4\beta -3\beta^2\right) + 
\left(1-8\beta+17\beta^2-12\beta^3\right)c + c^2 - c^3\right]
           \over 4 \left( 1 - 2 \beta  \right)^2 g^3}
\label{p1beta}
\end{equation}
and
\begin{eqnarray}
p_3 &=& {\beta ^3 \over 
     16 \left( 1 - 2 \beta  \right)^4 g^5}\Biggl[
     8 {{\left( 1 - 2 \beta  \right) }^4} 
    \left( 1 - 5 \beta  + 6 {{\beta }^2} - 2 {{\beta }^3} \right)  
\nonumber
\\
&&\hspace{0.5in} + 
   2 {{\left(1 - 2 \beta  \right) }^2} 
    \left( -9 + 77 \beta  - 209 {{\beta }^2} + 235 {{\beta }^3} - 
      90 {{\beta }^4} \right)  c 
\nonumber\\
&& \hspace{0.5in}+ 
   \left( 6 - 131 \beta  + 658 {{\beta }^2} - 1390 {{\beta }^3} + 
      1364 {{\beta }^4} - 531 {{\beta }^5} \right)  {c^2} 
\label{p3beta}\\
&&\hspace{0.5in} + 
   2 \left( 6 - 2 \beta  - 71 {{\beta }^2} + 135 {{\beta }^3} - 
      60 {{\beta }^4} \right)  {c^3} 
\nonumber\\
&&\hspace{0.5in} + 
   6 \left( -2 + 3 \beta  + 5 {{\beta }^2} - 9 {{\beta }^3} \right)  
    {c^4} + 6 \left( 1 -  \beta  \right)  {c^5} + 
   \left( -2 + \beta  \right)  {c^6} \Biggr]\ ,
 \nonumber
\end{eqnarray}
where $\beta$ satisfies the quintic equation
\begin{equation}
\beta^2 c \left(\beta  + c - 1 \right)^2 + \left( 1 - 2 \beta  \right)
   \left[\beta \left(\beta  + c - 1 \right) \left(c^2 -  
       \left( 1 - 2 \beta  \right)^2  \right) 
       - 2 g^2 \left( 1 - 2 \beta \right)\right]  = 0\ .
\label{beta}
\end{equation}
The relation
\begin{equation}
p_2 = {(1-c)\over g}p_1
\label{p2}
\end{equation}
gives $p_2$ in terms of $p_1$ \cite{cot1}.

Although the dual theory is directly related to the Ising theory
through the coordinate transformation (\ref{eq:transform}), the dual
theory is somewhat more difficult to solve  by matrix model
methods since the action is not of the form (\ref{eq:simpleaction}).
Nonetheless, because the action is quadratic in $Y$, this matrix can
be integrated out, allowing the model to be solved 
\cite{kostov,duk,ez}.                                       

In the dual theory, the correlation functions
\begin{equation}
\widehat{p}_n \equiv\frac{1}{ N} \left\langle{\rm Tr}\; X^n \right\rangle
\end{equation} 
correspond to the disk amplitude with boundary spins which are all
identical, but which may be of either spin.  Note that these
correlation functions also describe the Ising theory with free
boundary conditions, since $X$ corresponds to 
$(U+V)/\sqrt{2}$.  Inserting a $Y$ into the correlation function
corresponds to fixing the dual spins on the boundary to be of opposite
values on the two vertices separated by the edge labeled by $Y$.
Thus, each $Y$ corresponds to an insertion of the disorder operator.

The generating function                  
\begin{equation}
\widehat{\phi}(x)\equiv{1\over N}\left\langle{\rm Tr}\; 
{1\over 1-xX}\right\rangle = \sum_{k = 0}^{\infty} x^k \widehat{p}_k
\end{equation}
for the dual theory satisfies a cubic equation similar to
(\ref{eq:purecubic}) \cite{ez,cot1}.                          
This cubic equation is
\begin{equation}
\widehat{f}_3\widehat{\phi}^3 +
\widehat{f}_2\widehat{\phi}^2+\widehat{f}_1\widehat{\phi}+\widehat{f}_0 = 0 \ ,
\label{eq:dualcubic}
\end{equation}
with coefficients
\begin{equation}
\begin{array}{rcl}
\widehat{f}_0 &=& - 4c{\widehat{g}^2}(1+x\widehat{p}_1 +x^2\widehat{p}_2) +
2c\widehat{g}x(3-c)(1+\widehat{p}_1 x)  
- 2c{x^2}(1-c^2) - {\widehat{g}^2}{x^2}
\\
\widehat{f}_1 &=&  4c{\widehat{g}^2} - 2c\widehat{g}x(3-c) + 2c{x^2}(1-c^2) +
\widehat{g}{x^3}(1 - 3c) - 
      2{\widehat{g}^2}\widehat{p}_1{x^3}  
\\
\widehat{f}_2 &=& x^2\left({\widehat{g}^2} - \widehat{g}{x}(1-5c) -
      2c{x^2}(1+c)\right)
\\
\widehat{f}_3&=& \widehat{g}x^5\ .
\end{array}
\end{equation}
The coefficients $\widehat{p}_1,\widehat{p}_2$ can be related through
(\ref{eq:transform}) 
to the coefficients $p_i$ 
\begin{eqnarray}
\widehat{p}_1 & = &  \sqrt{2}p_1\\
\widehat{p}_2 & = &  {(1-c^2)p_1-g(1+gp_3)\over gc}\ . \nonumber
\end{eqnarray}

\section{Disk amplitudes}

We now take the cubic equations satisfied by the homogeneous disk 
amplitudes in the Ising and dual theories, and expand around the critical 
point to extract the continuum limits.

\subsection{Ising theory disk amplitude}

The critical values of $g$ and $c$ can be determined by finding the
radii of convergence for $\beta$.
They are given by \cite{kazakov,boukaz}
\begin{equation}
c_c=\frac{1}{ 27} {\left(-1 + 2 \sqrt{7} \right) }
\label{ccrit}
\end{equation}
and 
\begin{equation}
g_c={3^{-9/2}\sqrt{10}\left(- 1 + 2 \sqrt{7} \right)^{3\over 2}}\ .
\label{gcrit}
\end{equation}

{}From Eq. (\ref{beta}) it follows, using (\ref{ccrit}) and (\ref{gcrit}),
that
\begin{equation}
\beta_c=\frac{1}{9} ({{5 - {\sqrt{7}}}})\ ,
\label{betacrit}
\end{equation}
and therefore that
\begin{equation}
g_c p_{1c}= \frac{1}{5}  ({3 -\sqrt{7}})
\label{p1crit}
\end{equation}
and
\begin{equation}
g_c p_{3c}=\frac{1}{100}( {-699 + 40\cdot 7^{3\over 2}})\ .
\label{p3crit}
\end{equation}
We can now compute the critical expansion of $\beta$, $p_1$ and
$p_3$.  The expansion for $g$, written as
\begin{equation}
g=g_c \exp ({- \epsilon^2 t})\ ,
\label{eq:gexp}
\end{equation}
is expressed in terms of $\epsilon^2$, where the continuum limit 
corresponds to taking $\epsilon\rightarrow 0$, with $\epsilon$ scaling
as an inverse length.  This choice is natural since the parameter
$t$ is conjugate to area.  From this expansion it follows that
\begin{eqnarray}
gp_1&=&{\left( 3 - \sqrt{7} \right)\over 5}  
        \Biggl( 1 - {\left( 22 + 10 \sqrt{7} \right)\over 
            9}   \epsilon^{2} t
\nonumber\\
&& \hspace{0.5in}+ {5^{1\over 3} \left( 55 + 25 \sqrt{7} \right)  
              \over 36}\epsilon^{8/3} t^{4/3}
   + {5^{5\over 3} \left( 11 + 5 \sqrt{7} \right) \over 216} 
         \epsilon^{{10/3}} t^{5/3}  + {\cal O} (\epsilon^4t^{2})
\Biggr) 
\label{P1exp}
\end{eqnarray}
and
\begin{eqnarray}
gp_3&=& {-699 + 40 \cdot 7^{3\over 2}\over 100}  -
   {4 \left( 121 - 5 \cdot 7^{3\over 2} \right)\over 25} \epsilon^{2} t
\nonumber\\ 
&& \hspace{0.5in}+ {9 \cdot 5^{1\over 3} \over 2} \epsilon^{8/3} t^{4/3}
   + {3 \cdot 5^{2\over 3} \over 4}\epsilon^{{10/3}} t^{5/3}
+{\cal O} (\epsilon^4 t^{2})\ .
\label{P3exp}
\end{eqnarray}

Turning now to the expansion of $\phi$, we  take
\begin{equation}
u=u_c\exp ({- \epsilon z}) = g_c\left( 4 - \sqrt{7} \right)  
 \exp ({- \epsilon z}) \ ,
\label{vexp}
\end{equation}
so that $z$ is the boundary cosmological constant in the continuum
limit, conjugate to boundary length.  The critical value of ${u}$ is 
derived from the condition that
at ${u}={u}_c$ the cubic has repeated roots.  Taking the expansions 
(\ref{eq:gexp}), (\ref{P1exp}) and (\ref{P3exp}) for
$g$, $gp_1$ and $gp_3$ above and substituting into (\ref{eq:purecubic}),
we find \cite{staudacher,gn}
\begin{equation}
\phi = {3 \left( 2 + \sqrt{7} \right)\over 10}  
      \left( 1 - {\left( 1 + 2 \sqrt{7} \right)\over 3} \epsilon z + 
        \epsilon^{4/3} \phi_{(4/3)}(z,t) + \epsilon^{5/3} \phi_{(5/3)}(z,t)
+ {\cal O}(\epsilon^2)\right)  \ ,
\label{eq:diskexpansion}
\end{equation}
where $\phi_{(4/3)}(z,t)$ solves the equation
\begin{equation}
 27 \phi_{(4/3)}^3 - 9 \cdot  5^{4\over 3} \phi_{(4/3)} t^{4/3} - 50 t^{2}  +
   20 (2 + \sqrt{7})^2 t z^2 - (2 + \sqrt{7})^4 z^4 =0\ ,
\label{pp4eq}
\end{equation}
and $\phi_{(5/3)}(z,t)$ is given by
\begin{equation}
\phi_{(5/3)}(z,t) = {2 \phi_{(4/3)} (z,t)  \left(2+\sqrt{7}\right)\left(
        {z^2}\left(2+\sqrt{7}\right)^2-10 t\right)  z
        \over 27 \phi_{(4/3)}^2 (z,t) - 3 \cdot  5^{4\over 3} t^{4/3}}\ .
\label{pp5}
\end{equation}
The solution to (\ref{pp4eq}) is
\begin{equation}
\phi_{(4/3)}(z,t)={\left(2+\sqrt{7}\right)^{4\over 3}\over 3\cdot  
2^{4\over 3}} \Phi (z,T) \ ,
\end{equation}
where
\begin{equation}
T={20\over \left(2+\sqrt{7}\right)^2}t
\label{T}
\end{equation}
and
\begin{equation}
\Phi (z,T)\equiv\left[
\left(z+\sqrt{z^2-T}\right)^{4\over 3}
+\left(z-\sqrt{z^2-T}\right)^{4\over 3}
\right]\ .
\label{bigphi}
\end{equation}
As we will discuss, the universal part of $\phi$ is given by 
$\phi_{(4/3)}(z,t)$.  Up to rescaling, the relevant
behavior of this function is contained in
$\Phi(z,T)$.  The higher-order contribution $\phi_{(5/3)}(z,t)$
will be used to compute the two domain function in Section
5.1.

\subsection{Dual theory disk amplitude}

{}From the dual cubic (\ref{eq:dualcubic}) and the relation
$\widehat{g}=g/\sqrt{2}$, we can use (\ref{ccrit}), (\ref{eq:gexp}),
(\ref{P1exp}) and (\ref{P3exp}), with
\begin{equation}
{x}={g_c\left(13-\sqrt{7}\right)
\over 6\sqrt{2}} \exp ({-\epsilon z})\ ,
\label{vhatexp}
\end{equation}
to obtain the critical expansion for $\widehat{\phi}$.  
We find \cite{ez}                                       
\begin{equation}
  \widehat{\phi}
  ={2\left(1+\sqrt{7}\right)\over 5}\left(1-\sqrt{7} \epsilon^{} z
  + \epsilon^{4/3} \widehat{\phi}_{(4/3)}(z,t) 
  + \epsilon^{5/3} \widehat{\phi}_{(5/3)}(z,t)
  + {\cal O}(\epsilon^2)\right)\ ,
  \label{eq:dualexpansion}
\end{equation}
with
\begin{equation}
\widehat{\phi}_{(4/3)}(z,t)={\left(1+\sqrt{7}\right)^{4\over 3}\over
2^{5/3}}\Phi (z, \widehat{T})
\label{pp4hat}
\end{equation}
and 
\begin{equation}
\widehat{\phi}_{(5/3)}(z,t)=0\ ,
\label{pp5hat}
\end{equation}
where
\begin{equation}
\widehat{T}={10\over \left(1+ \sqrt{7}\right)^2} t\ .
\label{That}
\end{equation}
Notice that $\widehat{\phi}_{(4/3)}$, which expresses the 
universal behavior in the continuum limit, involves the same function 
$\Phi$ (given by (\ref{bigphi})) which arose in the original Ising theory.

\subsection{Comparison with Liouville theory}

To compare the universal scaling behavior of the disk amplitude in the
Ising and dual theories, we recall some of the basic results of Liouville
theory.  These results were first given in \cite{kpz,david,dk,mss}; for a
review see \cite{mg}.  In Liouville theory the scaling behavior of the
partition function $Z (A)$ on a sphere of fixed area $A$ is
\begin{equation}
Z (A) = A^{{\Gamma}-3}\ ,
\label{eq:partitionprediction}
\end{equation}
where for the $c = 1/2$ model, $\Gamma=-1/3$.  On a disk with area $A$ and
boundary length $l$, the partition function  for conformally
invariant boundary conditions scales as
\begin{equation}
Z (l,A) = l^{ -3 + Q/\gamma} A^{{ -Q}/\gamma} e^{-l^2/A}
=l^{ -2/3} A^{{ - 7}/3} e^{-l^2/A}\ ,
\label{eq:diskprediction}
\end{equation}
where $Q = 7/\sqrt{6}$, $\gamma = \sqrt{3}/\sqrt{2}$.

The conceptually simplest way to compare the predictions of Liouville
theory to the matrix model is to directly translate
(\ref{eq:partitionprediction}) and (\ref{eq:diskprediction}) into
statements about the asymptotic behavior of coefficients in the matrix
model generating functions.  For example, consider expanding the matrix
model partition function $Z (g)$ as a function of $g$ at $c = c_c$:
\begin{equation}
Z (g) = \sum_{n = 0}^{ \infty}  z_ng^n\ .
\end{equation}
Since $n$ is proportional to the area of the surface, to recover
(\ref{eq:partitionprediction}) in the continuum we require that
$z_n \rightarrow g_c^{-n} n^{-10/3}$ asymptotically as $n
\rightarrow \infty$.  As another example, consider $p_3$; this is 
the partition function for a disk with three boundary edges, which
may be thought of as a sphere with a single triangle marked.
We can expand
\begin{equation}
p_3 (g) = \sum_{n} m_n g^n\ .
\end{equation}
The coefficients $m_n$ essentially count the total Boltzmann weight of all
triangulations with $n + 1$ triangles, with one triangle marked, and
should scale as
\begin{equation}
m_n \rightarrow n \cdot g_c^{-n} n^{-10/3}= g_c^{-n} n^{-7/3}\ .
\end{equation}
This scaling behavior can be verified numerically by expansion of the 
exact solution (\ref{p3beta}) to high order in $g$.

An advantage of this way of comparing matrix model results with the
predictions of Liouville theory is that there is no possible confusion
in extracting ``universal'' parts of the matrix model results.  The
usual approach to this comparison, however, is to take the Laplace
transform of (\ref{eq:partitionprediction}) and
(\ref{eq:diskprediction}).  This approach has the advantage that the
calculations can be performed analytically; however, we now must
be careful in extracting the portion of the matrix model result which
actually contains information about the asymptotic form of the
generating function coefficients.  A rule of thumb which is usually
reliable is that the first nonanalytic term in the expansion
of an amplitude around $\epsilon = 0$ gives enough information to find 
the asymptotic form of the coefficients and to check agreement with 
Liouville theory. Let us consider a simple example.  It is not hard to 
verify that when $c_n \rightarrow n^{s/t-1}$ as $n \rightarrow \infty$, 
with $s,t$ relatively prime integers, we have
\begin{equation}
\sum_{n} c_n e^{-\epsilon n} = f (\epsilon) + \Gamma(s/t)\epsilon^{-s/t}
+ {\cal O}(\epsilon^{-(s -1)/t})\ , 
\label{eq:example}
\end{equation}
where $f(\epsilon)$ is analytic in $\epsilon$.  The function $f(\epsilon)$
is referred to as ``nonuniversal'', since it is not determined exclusively 
by the asymptotic behavior of $c_n$, but depends also on the specific values 
of the $c_n$'s for small $n$.  The first nonanalytic term, on the other
hand, is determined only by the asymptotic behavior of $c_n$, and is thus
considered ``universal''.   Thus the rule of thumb works in this simple
example (although it would fail if $s/t$ were an integer).   
In Section \ref{sec:multiple}, however, we will find that in
some situations a more careful analysis is required.

We now consider the disk amplitude.  Writing $p_k$ as a power series in $g$
\begin{equation}
p_k =\sum_{n = 0}^{ \infty}  g^n q_{k,n}\ ,
\end{equation}
in order to recover the Liouville prediction (\ref{eq:diskprediction})
we require asymptotic behavior
\begin{equation}
q_{k,n} \rightarrow g_c^{-n}u_c^{-k}k^{1/3} n^{-7/3}
e^{-k^2/n}\ .
\end{equation}
(Note that we have an extra power of $k$ since one edge is marked.)
To see that this prediction is satisfied, we take the discrete Laplace
transform
\begin{equation}
\sum_{n,k = 0}^{\infty} 
e^{-\epsilon kz-\epsilon^2 nt} k^{ 1/3}  n^{{ -7}/3} e^{-k^2/n}
=  f (\epsilon z,\epsilon^2 t) + 2^{-4/3}\Gamma (4/3)\Gamma (-4/3)
\epsilon^{4/3}\Phi (z,t/4) + {\cal O}(\epsilon^{5/3})\ ,
\end{equation}
where $f(\epsilon z,\epsilon^2 t)$ is an analytic function of $\epsilon$.
As in the example (\ref{eq:example}), the universal behavior is contained
in the first nonanalytic term, which is given in terms of scaling
variables $(z,t/4)$ by the function $\Phi$ familiar from
the disk amplitudes previously discussed. 
Thus we see that the expansions (\ref{eq:diskexpansion}) and
(\ref{eq:dualexpansion}) of the disk amplitudes $\phi$ and $\widehat{\phi}$
both have universal behavior which agrees with the prediction of Liouville
theory for the $c = 1/2$ model.

\section{Spin/Disorder operators on the sphere}

The scaling dimension of the disorder operator in the dual theory was
computed in \cite{kostov} and shown to agree with the KPZ/DDK Liouville
results.  In this section we compute an exact expression for 
correlation functions of two disorder operators on the sphere in
both the Ising and the dual theories.  We show that the results agree and
give the correct scaling dimension to the disorder operator.  This provides
evidence that  Kramers-Wannier duality is maintained for operators in the
bulk after coupling to gravity.

We compute the correlation function of two disorder operators on the
sphere by considering a cylinder with a single disorder operator on each
boundary; taking the boundaries to have a length which vanishes in the
scaling limit, we obtain the correlation function of two disorder operators
on the sphere.  Notice that, in 
order to consider boundaries with a single disorder operator, it is
necessary to introduce a twist in the ${\bf Z}_2$ bundle over
the cylinder of which the spin configuration forms a section.  (Even so,
the total number of disorder operators on the entire surface must be
even.)  There is no way to introduce such a twist if the
fundamental variables simply represent up and down spins; instead we
are forced to use variables living between the spin sites, which indicate
whether the spins on either side are like or unlike.  Such variables are
natural in the dual theory, but are incompatible with the formulation of
the original Ising theory as a 2-matrix model.  In the calculation below,
therefore, we use an alternate set of variables introduced in \cite{cot1}, 
in which boundary data are given in terms of spin correlations.

\subsection{Disorder operators in the dual theory}

We begin by looking at the correlation function of two spin operators in
the Ising theory or equivalently at the correlation function of two
disorder operators in the dual theory,
\begin{equation}
\widehat{\omega}_0 \equiv
\frac{1}{2} \left\langle {\rm Tr}\; (U -V) \; \cdot \;{\rm Tr}\; 
(U -V)\right\rangle
= \left\langle {\rm Tr}\; Y \; \cdot \;{\rm Tr}\; Y \right\rangle\ .
\end{equation}
Geometrically, this corresponds to a cylinder for which both boundaries
are a single $Y$ edge (equivalent in the continuum limit to a sphere
with two marked points where disorder operators are inserted).

To compute $\widehat{\omega}_0$, we find an equation satisfied by 
the generating function
\begin{equation}
\widehat{\omega}(x)  \equiv
\sum_{k= 0}^{\infty}\left\langle {\rm Tr}\; (Y X^k) \; \cdot \;
{\rm Tr}\; Y \right\rangle x^k
\label{eq:yxky}
\end{equation}
in the dual model.  Using the formalism of \cite{cot1}, an equation
for this generating function may be derived by considering the result
of removing a single $Y$ edge from triangulations contributing to
(\ref{eq:yxky}).  We obtain
\begin{equation}
x\widehat{\omega} (x) = \frac{1}{1 + c}  \left[ x\phi (x) + 2 \widehat{g} 
( \widehat{\omega}
(x) -\widehat{\omega}_0) 
\right]\ ,
\end{equation}
so
\begin{equation}
(x (1 + c)-2\widehat{g}) \widehat{\omega} (x) = x \phi (x) -2\widehat{g} 
\widehat{\omega}_0\ .
\end{equation}
The left hand side of this equation vanishes near $x = \widehat{g} = 0$ along
the curve $x = 2\widehat{g}/(1 + c)$.  It  follows that
\begin{equation}
\widehat{\omega}_0 (\widehat{g}) = \frac{1}{1 + c}  
\widehat{\phi} \left(\frac{2\widehat{g}}{1 + c} \right).
\end{equation}
At the critical point, we have
\begin{equation}
x_c = \frac{2\widehat{g}_c}{1 + c_c} \ .
\end{equation}
The expansion for $ \widehat{\phi}
({2\widehat{g}}/(1 + c))$ is just the expansion for $\widehat{\phi} (x)$ 
given in (\ref{eq:dualexpansion})
with $z =
0$.  Thus, we  have
\begin{equation}
\widehat{\omega}_0 = \frac{1 + 2 \sqrt{7}}{5} 
\left( 1 -  \frac{5^{2/3}}{2}  \epsilon^{4/3}t^{2/3}  + 
{\cal O} (\epsilon^2 t )\right)\ .
\label{eq:expandtheta}
\end{equation}
Below we will see that the universal behavior, represented by the
non-analytic term of order $\epsilon^{4/3}$, agrees with the prediction
of Liouville theory.

\subsection{Disorder operators in the Ising theory}

The correlation function of two disorder operators in the Ising theory, or
equivalently of two spin operators in the dual theory, cannot be expressed
in terms of expectation values of the matrices $U$ and $V$ or $X$ and
$Y$. However, it turns out that we can do the computation using an
alternative description of the Ising theory, introduced in
Ref. \cite{cot1}, in which the boundary conditions are specified by
labeling the vertices between boundary edges as either ``stick'' or
``flip'', depending on whether the spins on either side are like or
unlike.  These variables allow us to include configurations with
boundary conditions which do not appear in the matrix model formulation.

In Ref. \cite{cot1}, we derived a generating equation for the stick/flip
generating function in terms of free variables, and this was used to obtain
a cubic equation for the homogeneous disk amplitude (all sticks on the
boundary) that is identical to equation (\ref{eq:purecubic}).
In a similar way, we can write an equation for the generating
function
\begin{equation}
\omega(s)\equiv \sum_{s=0}^\infty\omega_k s^k\ ,
\end{equation}
where $\omega_k$ is the weight associated with a surface with two
boundaries, one with a single vertex marked as a flip, and the other with
$k$ stick vertices and one flip vertex.  Such a configuration involves
a single twist in the ${\bf Z}_2$ bundle over the cylinder, and cannot
be described as a correlation function of the matrix model operators $U$ 
and $V$.  The equation for $\omega(s)$ reads (following the conventions of
Ref. \cite{cot1})
\begin{equation}
\begin{array}{rcl}
s\omega(s)&=&\displaystyle{1\over 2(1-c)}\Bigl[ 2s^3\omega(s) + 
s^3\omega(s)(\phi_-(s)-1)
\\
&&\qquad\qquad
\displaystyle
+ 2g(\omega(s)-\omega_0-s\omega_1)+2gs\omega_1+2s(\phi_-(s)+1)\Bigr]\ ,
\end{array}\label{eq:ido}
\end{equation}
where $\phi_-(s)$ is the homogeneous disk amplitude for a boundary of all
stick vertices, and is equal to $2\phi(s)-1$ where $\phi(s)$ is the
homogeneous disk amplitude (\ref{eq:phidef}) of the Ising theory.
Therefore (\ref{eq:ido}) becomes
\begin{equation}
\omega(s)\left[(1-c)s-s^3\phi(s)-g\right]=2s\phi(s)-g\omega_0\ .
\label{eq:ido2}
\end{equation}
As above, it follows that 
\begin{equation}
  \omega_0(g)={2\bar{s}\phi(\bar{s})\over g}
  ={2(1-c)\bar{s}-2g\over g\bar{s}^2}\ ,
\end{equation}
where $\bar{s}(g)$ is defined by the equation
\begin{equation}
(1-c)\bar{s}-\bar{s}^3\phi(\bar{s})-g=0	  \ .
\label{eq:sbar}
\end{equation}
At the critical point,
\begin{equation}
(1-c)s_c-{s_c}^3\phi_c-g_c=0\ ,
\end{equation}
where $s_c=(4-\sqrt{7})g_c$ as in (\ref{vexp}). The expansion for 
$\bar{s}$ can be derived by expanding (\ref{eq:sbar}),
\begin{equation}
\bar{s}=(4-\sqrt{7})g_c\left(
1+ {5^{2/3}(2-\sqrt{7})\over 6}\epsilon^{4/3}t^{2/3}
+ {\cal O} (\epsilon^2 t)
\right) \ ,
\end{equation}
from which it follows that
\begin{equation}
\omega_0={3(1+2 \sqrt{7})\over 5}\left(
1- {2 \,\cdot\,5^{2/3}\over 3}  \epsilon^{4/3}t^{2/3}
+ {\cal O}(\epsilon^2t)\right)\ .
\label{eq:expandtheta2}
\end{equation}

\subsection{Comparison with Liouville theory}

We can now compare the above results to the predictions of KPZ/DDK based on
Liouville theory \cite{kpz,david,dk}.  The scaling dimension of the disorder 
and spin operators in flat space is $\Delta_0 = 1/16$.  The two point function
of the gravitationally dressed operators on a sphere of fixed area $A$ has
scaling behavior
\begin{equation}
\omega_{L} (A) \sim A^{2-2 \Delta} Z (A)\ ,
\end{equation}
where $\Delta = 1/6$ is computed by
\begin{equation}
\Delta = 1-\frac{\alpha}{\gamma}
\end{equation}
with  $\alpha$ satisfying
\begin{equation}
\Delta_{0} = 1 + \frac{1}{2} \alpha (\alpha-Q)\ .
\label{eq:bulkscaling}
\end{equation}
Thus, we expect
$\omega_{L} (A) \sim A^{-5/3} $.
For the matrix model to reproduce this behavior, we would expect that
at $c = c_c$ the expansion of $\omega_0 (\widehat{g})$
\begin{equation}
\omega_0 (\widehat{g}) = w_n \widehat{g}^n
\end{equation}
would have coefficients $w_n$ with asymptotic behavior
\begin{equation}
w_n \rightarrow \widehat{g}_c^{-n} n^{-5/3}\ .
\end{equation}
Taking $\widehat{g} = \widehat{g}_c \exp ({-t\epsilon^2})$ as in
(\ref{eq:gexp}), we have the 
summation 
\begin{equation}
\omega_0 (\widehat{g}) =\sum_{n} w_n \widehat{g}_c^n \exp ({-nt \epsilon^2})
=\omega_c  +  \omega_{(4/3)} \epsilon^{4/3}t^{2/3}+ {\cal O}(\epsilon^2 t)\ ,
\end{equation}
where $\omega_{(4/3)}$ is a constant coefficient.  This agrees
precisely with the universal terms in (\ref{eq:expandtheta}) and
(\ref{eq:expandtheta2}), verifying that the continuum limits of
both discrete models agree with the predictions of Liouville theory.
The agreement between the Ising and dual theories 
is evidence that Kramers-Wannier duality, which exchanges the
gravitationally dressed spin and disorder operators in the bulk, is a 
symmetry of the $c = 1/2$ theory even after coupling to gravity.

\section{Disorder operators on the disk}
\label{sec:multiple}

We now proceed to derive analytic expressions for the correlation
functions of multiple disorder operators on the disk, both in the
regular and dual theories.  We begin by using loop equation methods to
write the exact expressions for the two operator correlation
functions.  We then compute the continuum limit of these expressions, and
compare  the results to Liouville theory.  

\subsection{Ising theory}

The generating function for the two domain disk amplitude in the Ising
theory, corresponding to the correlation function of two disorder operators 
which separate the domains, is defined by
\begin{equation}
\sigma (u,v) \equiv \frac{1}{ N} 
\left\langle  {\rm Tr}\; \left(\frac{ uU}{1-uU} \cdot
 \frac{vV}{1-vV}  \right) \right\rangle= \sum_{k, l = 1}^{ \infty}
s_{k, l}u^k v^l\ .
\end{equation}
Using the techniques of \cite{cot1}, one can derive the following
equation for $\sigma(u,v)$ (essentially the same equation was derived
in \cite{staudacher})
\begin{eqnarray}
\sigma(u,v)& =&
     -\biggl({c g u^2 
     + g v^2 -
     (1 - {c^2}) u v^2+
     u^2 v^2\Bigl[u\phi(u)
     +cv\phi(v)\Bigr]}\biggr)^{-1}
 \nonumber\\
&&\hspace{0.2in}\times\biggl(c g u^2 \left(\phi (v)-1\right) 
     + g v^2 \left(\phi (u) -1\right)
     - (1-c^2) uv^2\left(\phi (u)-1\right)
\nonumber\\
&&\hspace{0.5in} -  uv p_1\left(v+cu\right)
     - g u v \Bigl[cu\phi_v(u) 
     +v \phi_v(v)\Bigr]
\label{sigmaeq}\\
&&\hspace{0.5in}   +  u^2 v^2 \left[
     \Bigl(u \phi (u)+cv\phi (v)\Bigr)
     \left(\phi(u)+\phi(v)-1\right) \right]\biggr)\ ,
\nonumber
\end{eqnarray}
where
\begin{equation}
\phi_v(u)\equiv
{1\over N}\sum_{n=1}^\infty u^n \left\langle\Tr V U^n\right\rangle
\end{equation}
is related to $\phi(u)$ and $p_1$ by
\begin{equation}
\phi_v(u) = {1\over  cu^2}\left[u(g - c u) p_1 -
      u^3\left(2-\phi(u)\right)^2
      - (g-u)(\phi (u)  -1)\right]\ .
\label{sigma1}
\end{equation}
Expanding in $\epsilon$ and using  separate cosmological constants for
the two boundary variables $u$ and $v$,
\begin{eqnarray}
u&=&g_c\left( 4 - \sqrt{7} \right)  
   \exp ({-\epsilon^{} z_u})
\nonumber\\
v&=&g_c\left( 4 - \sqrt{7} \right)  
\exp ({- \epsilon^{} z_v})\ ,
\label{qrexp}
\end{eqnarray}
\newpage
\noindent 
we get
\begin{eqnarray}
\sigma&=&\frac{1}{ 10}  \left[  {(9 - 2 \sqrt{7})}  -
   {\left( 23 + 10 \sqrt{7} \right)  \epsilon^{} 
       \left( z_u + z_v \right) }  \right.
\nonumber\\
&&\hspace{0.5in} \left. + 
   {9 \left(3 + \sqrt{7} \right)  \epsilon^{4/3} 
       \left( \phi_{(4/3)}(z_u,t) + \phi_{(4/3)}(z_v,t) \right) }  \right.
\nonumber\\
&&\hspace{0.5in} +
   {9 \left( 3 + \sqrt{7} \right)  \epsilon^{5/3}
       \left( \phi_{(5/3)}(z_u,t) + \phi_{(5/3)}(z_v,t) \right) }
\label{sigmaexp}\\&&\hspace{0.5in}
\left. +2^{-8/3} \left( 2 + \sqrt{7} \right)^{11/3}  \epsilon^{5/3} 
\Sigma (z_u,z_v,T)+ {\cal O}(\epsilon^2)\right]\ ,
 \nonumber
\end{eqnarray}
where $T$ is given by (\ref{T}) and the function $\Sigma$ is given by
\begin{equation}
\Sigma (z,y,T)  \equiv
\frac{3T^{4/3}- \left[\Phi (z,T)^2 +\Phi (y,T) \Phi (z,T) + 
\Phi (y,T)^2\right]}{z + y}\ .
\label{eq:sigmadefinition}
\end{equation}
We shall see below that $\Sigma (z,y,T)$ plays a similar role for the
two-domain amplitude that $\Phi(z,T)$ played for the one-domain
amplitude, capturing the universal behavior in the continuum limit.

\subsection{Dual theory}

In the dual theory, the two-disorder-operator correlation function is
given by
\begin{equation}
\widehat{\sigma} (x,\w)\equiv
 \frac{1}{N} \left\langle  {\rm Tr}\; \left(\frac{ 1}{1- xX} \cdot Y
 \frac{1}{1- wX} Y \right) \right\rangle =\sum_{k, l =  0}^{ \infty}
\widehat{s}_{k, l}u^k v^l\ .
\end{equation}
Using the method of \cite{cot1} it can be shown that $\widehat{\sigma}$
satisfies the equation
\begin{eqnarray}
\widehat{\sigma}(x,\w)&=& \biggl( x \w 
      \left( -\widehat{g} x - \widehat{g} \w +
      (1+c)x \w \right)\biggr)^{-1}
   \nonumber\\
  &&\hspace{0.2in}\times
      \biggl(-\widehat{g}x - \widehat{g}\w + 
      2 (1-c) x \w 
      - 2 \widehat{g}\widehat{p}_1 x \w +
      x^2 \w^2 \widehat{\phi}
      (x) \widehat{\phi} (\w)
\label{sigmahateq}\\
&&\hspace{0.5in} + \widehat{g} \w \widehat{\phi} (x) 
   - (1-c)  x \w \widehat{\phi}
      (x) +
       x^3 \w \widehat{\phi} (x)^2
\nonumber\\
&&\hspace{0.5in} + \widehat{g} x \widehat{\phi} (\w) 
   - (1-c) x \w \widehat{\phi}
      (\w) +
      x \w^3 \widehat{\phi} 
      (\w)^2\biggr)\ .
\nonumber
\end{eqnarray}
As in the previous section, we expand with different cosmological
constants for each domain
\begin{eqnarray}
x&=&{g_c\left(13-\sqrt{7}\right)\over 
6\sqrt{2}}
\exp ({-\epsilon z_x})
\nonumber\\
\w&=&{g_c\left(13-\sqrt{7}\right)\over 
6\sqrt{2}}
\exp ({-\epsilon z_\w})\ .
\label{qrhatexp}
\end{eqnarray}
The critical expansion of $\widehat{\sigma}(x,\w)$ is then
\begin{eqnarray}
\widehat{\sigma} &=&   \frac{1}{25}  \left[
{18 \left( 2+\sqrt{7}\right)}  -
   {18 \left(16 + 5 \sqrt{7} \right)  {\epsilon^{}} 
       \left( z_x + z_\w \right)} \right.
\nonumber\\&&\hspace{0.5in}
 + {72 \left(2 + \sqrt{7} \right)  {\epsilon^{4/3}} 
       \left( \widehat{\phi}_{(4/3)}(z_x,t) + \widehat{\phi}_{(4/3)}
    (z_\w,t) \right)}  
\label{sigmahatexp}\\&&\hspace{0.5in} 
\left. +3 \cdot 2^{-4/3} \left(5+\sqrt{7}\right)  
\left( 1 + \sqrt{7}\right)^{8/3}{\epsilon^{5/3}} 
\Sigma (z_x,z_\w, \widehat{T})+ {\cal O}(\epsilon^2)\right]\  \nonumber
\end{eqnarray}
where $\Sigma (z_x,z_w, \widehat{T})$ is again given by
(\ref{eq:sigmadefinition}).  The reappearance of this function 
is an indication of duality, as we explain below.

\subsection{Comparison with Liouville theory}

To compare these results with Liouville theory, we recall the work of
Martinec, Moore, and Seiberg \cite{mms} in which the scaling
dimensions of boundary operators in Liouville theory were derived
using methods analogous to those of KPZ/DDK.  According to their
analysis, the scaling dimension $\Delta_b$ of a boundary operator with
flat space scaling dimension $\Delta_{b0}$ is given by
\begin{equation}
\Delta_b = 1-\frac{\alpha_b}{\gamma}\ ,
\end{equation}
where 
\begin{equation}
\Delta_{b0} = 1 + \alpha_b (2 \alpha_b-Q)\ .
\label{eq:boundaryscaling}
\end{equation}
The flat space scaling dimension of the spin and disorder operators 
was shown by Cardy to be $\Delta_{b0}= 1/2$.  Thus, we have
\begin{equation}
\Delta_b = 2/3\ .
\end{equation}
It follows that the correlation function of two disorder operators on
the boundary of a disk with area $A$ and boundary length $l$ should be
of the form 
\begin{equation}
R(l',l -l',A) = l^{-2/3 + (1-2\Delta_b)} A^{{- 7}/3}
r\left(\frac{l'}{ l-l'}, \frac{l^2}{A}\right)\ ,
\label{eq:continuumcorrelation}
\end{equation}
where $l'$ and $l-l'$ are the boundary lengths separating the
operators,  and  $ r$ is an undetermined function of the dimensionless
ratios.  

Some care is needed to compare (\ref{eq:continuumcorrelation}) with the
matrix model results.  From the scaling dimension of $R$ we see that the
universal term in $\sigma$ should scale as $\epsilon^{5/3}$.  However, the
leading nonanalytic term in $\sigma$ is of order $\epsilon^{4/3}$.  In fact,
this leading term is not universal, but arises from contact terms where the
distance between the two disorder operators vanishes in the scaling
limit.  An easy way to see this heuristically is that the universal
behavior of $\sigma$ should be unchanged if we subtract from it only those
terms where the two disorder operators are adjacent.  This corresponds to
taking $\sigma (u, v)\rightarrow \sigma (u, v)-\phi_v(u)-\phi_v(v)$. 
The expansion of $\phi_v$ is essentially the same as
the expansion of $\phi$, so the effect of this modification is to
change the coefficients in front of $\phi_{(4/3)}$ and $\phi_{(5/3)}$ in
the expansion (\ref{sigmaexp}) of $\sigma$.

This suggests that the only universal terms in $\sigma$ and
$\widehat{\sigma}$ are those which are singular in  {\it both} boundary
cosmological constants.  We now perform a simple calculation which 
serves as a more concrete demonstration of this fact.

Integrating out $A$ from (\ref{eq:continuumcorrelation}), with a
vanishing bulk cosmological constant, we have
\begin{equation}
S (l',l-l') =l^{-11/3}  s\left(\frac{l'}{ l-l'}\right)\ ,
\label{eq:liouvillecorrelator}
\end{equation}
where $s$ is an undetermined function.  The simplest function with the
correct scaling and symmetry properties is
\begin{equation}
\widetilde{S} (l', l-l') = l^{-7/3} (l'^{-4/3}+ (l-l')^{-4/3})\ .
\end{equation}
Imagine that the matrix model exactly reproduced this asymptotic
behavior; we would then have
\begin{equation}
\widetilde{s}_{k, l} \sim u_c^{-k-l} (k + l)^{-7/3}
\left( k^{-4/3} + l^{-4/3} \right).
\label{eq:approximateform}
\end{equation}
Summing over all boundary lengths, this would give
\begin{eqnarray}
\widetilde{\sigma}  & = &  \sum_{k, l = 1}^{ \infty} 
e^{-\epsilon (kz_u + lz_v)} (k + l)^{-7/3}
\left( k^{-4/3} + l^{-4/3} \right)\nonumber\\
 & = & \widetilde{\sigma}_c + \widetilde{\sigma}_{(1)} \epsilon (z_u + z_v)
+ \widetilde{\sigma}_{(4/3)}\epsilon^{4/3}(z_u^{4/3}+z_v^{4/3})
\\ & &\hspace{0.5in}
+ \widetilde{\sigma}_{(5/3)}\epsilon^{5/3} (z_u^{5/3}+z_v^{5/3})
+ \widetilde{\Sigma}(z_u, z_v) \epsilon^{5/3}+{\cal O} (\epsilon^2)\ ,
\nonumber
\end{eqnarray}
where $\widetilde{\sigma}_{(1)},\widetilde{\sigma}_{(4/3)}, 
\widetilde{\sigma}_{(5/3)}$
are nonzero constants, and $\widetilde{\Sigma} (z_u, z_v)$ is a linear
combination of hypergeometric functions.  In this case, even though the
coefficients $\widetilde{s}_{k,l}$ agree with the Liouville theory predictions,
extra nonuniversal terms are generated which are not simultaneously 
singular in $z_u$ and $z_v$.  Thus, in this example, the universal behavior
is described by the leading term which is nonanalytic in both
$z_u$ and $z_v$. 

The hypergeometric function $\widetilde{\Sigma}$ is not
precisely equal to the universal form $\Sigma$, indicating that the exact
coefficients $s_{k,l}$ are more complicated than $\widetilde{s}_{k,l}$.
However, from the discussion above we see that the universal parts of the
expansions (\ref{sigmaexp}) and (\ref{sigmahatexp}) are contained in
the terms proportional to $\Sigma$, which are singular in both
boundary cosmological constants.  Because these terms take an
identical form, we see that the correlation function of two boundary
disorder operators is identical in the Ising and dual theories.  This
result clears up the puzzle from \cite{staudacher}, and gives
strong evidence that Kramers-Wannier duality is preserved in the
presence of gravity, even on an open string with nontrivial boundary
conditions. 

\section{Conclusion}

We have calculated the exact two-point functions for spin and
disorder operators on the sphere and on the boundary of a disk in the
$c = 1/2$ string, both in the Ising and dual Ising matrix model
formulations.  We find that the universal parts of the
correlation functions are equal in
both models, suggesting that the Kramers-Wannier duality symmetry of
the $c = 1/2$ conformal field theory is also a symmetry of the full $c
= 1/2$ string.

In the $c = 1/2$ CFT, Kramers-Wannier duality is broken by the GSO
projection operator \cite{ginsparg} which divides the full Hilbert
space into two parts, corresponding to the two dual theories. 
Presumably an analogous procedure is at work
in the gravitationally coupled theory, although it is unclear how
to describe such a mechanism from the point of view of the 
discretized theory.

Since the duality of a single Ising spin seems to be preserved under
coupling to gravity, it is interesting to consider the result of
coupling two Ising spins to gravity.  It is known that the CFT
describing two Ising spins is a $c = 1$ orbifold theory with $r = 1$
\cite{ginsparg}.  The Kramers-Wannier duality of the two Ising spins
in this theory gives a ${\bf Z}_2\times {\bf Z}_2$ symmetry to the
Hilbert space.  It would be interesting to understand the relationship
of this duality symmetry to the usual $c = 1$ duality.

It is straightforward to generalize the methods of this paper to
calculate correlation functions between arbitrary numbers of disorder
and spin operators in either the Ising or dual theory.  (A calculation
along these lines was done in \cite{staudacher}).  For a correlation
function of $2k$ disorder operators on the boundary of a disk, the
Liouville prediction is that the correlation function should scale as
$7/3-2k/3$.  Using the methods of \cite{cot1}, it is also possible to
calculate these correlation functions in the presence of a boundary
magnetic field; this work is currently in progress.

\section*{Acknowledgments} 
We would like to thank M. Douglas, N.
Elkies, J. Goldstone, M. Li, A.  Matytsin, G. Moore, and M. Staudacher
for helpful conversations.  This work was
supported in part by the National Science Foundation under grants
PHY/9200687 and PHY/9108311, by the U.S. Department of Energy (D.O.E.)
under cooperative agreement DE-FC02-94ER40818, by the divisions of 
applied mathematics of the D.O.E. under contracts DE-FG02-88ER25065
and DE-FG02-88ER25065, and by the European Community
Human Capital Mobility programme.

\end{document}